\documentclass[12pt]{article}
\usepackage{epsfig}
\usepackage{amsfonts}
\usepackage{latexsym}
\usepackage{amsmath,amssymb}
\usepackage{mathrsfs}
\usepackage{hyperref}
\usepackage{setspace}
\usepackage{color}
\usepackage{bm}
\usepackage{slashed}
\usepackage{braket}
\numberwithin{equation}{section}

\begin{document}

\begin{titlepage}
\unitlength = 1mm


{\hfill YITP-22-55}
\vskip 1cm
\begin{center}

{\large {\textsc{\textbf{Graviton non-gaussianity in $\alpha$-vacuum}}}}

\vspace{1.8cm}
Sugumi Kanno$^a$ and Misao Sasaki$^{b,c,d}$

\vspace{1cm}

\shortstack[l]
{\it $^{a}$Department of Physics, Kyushu University, Fukuoka 819-0395, Japan
}\\
{\it $^{b}$Kavli Institute for the Physics and Mathematics of the Universe (WPI), \\
	University of Tokyo, Chiba 277-8583, Japan}\\
{\it $^{c}$Center for Gravitational Physics,
Yukawa Institute for Theoretical Physics,\\
Kyoto University, Kyoto 606-8502, Japan}
\\
{\it $^{d}$Leung Center for Cosmology and Particle Astrophysics,\\
National Taiwan University, Taipei 10617, Taiwan
}

\vskip 1.5cm

\begin{abstract}
\baselineskip=6mm
We compute the leading order non-Gaussianity, i.e., the bispectrum, of the tensor perturbation in the general $\alpha$-vacuum on de Sitter space in general relativity. In addition to the well-known Bunch-Davies (BD) vacuum, there exits an infinite number of de Sitter invariant vacua represented by a real parameter $\alpha$ and a phase $\phi$, with $\alpha=0$ being the BD vacuum. They are called $\alpha$-vacua. In the standard slow-roll inflation, as de Sitter invariance no longer applies, the $\alpha$-vacua lose its relevance in the rigorous sense. Nevertheless, if we assume that the parameter $\alpha$ is only weakly dependent on the wavenumber with an appropriate UV cutoff, we may consider pseudo-$\alpha$-vacua. In the case of false vacuum inflation where the background spacetime is pure de Sitter, a non-trivial (non-BD) $\alpha$-vacuum could indeed be realized.
We find an intriguing result that the bispectrum may be exponentially enhanced to be detectable by observation even if the spectrum is too small to be detected.
\end{abstract}

\vspace{1.0cm}

\end{center}
\end{titlepage}

\pagestyle{plain}
\setcounter{page}{1}
\newcounter{bean}
\baselineskip18pt

\setcounter{tocdepth}{2}

\tableofcontents

\section{Introduction}

Triggered by the discovery of gravitational waves (GWs) from a merging black hole binary by LIGO~\cite{LIGOScientific:2016aoc} and subsequent discoveries of many merging compact binaries, GWs have become one of the hot topics in astrophysics and cosmology. In particular, a lot of attention has been paid to possible GW signatures from the early universe in the last few years. Among them, one of the important sources of GWs is the one that comes from vacuum fluctuations during inflation~\cite{Kawamura:2011zz, AmaroSeoane:2012km}.

Conventionally the tensor mode (graviton) vacuum fluctuations are considered to be in the adiabatic vacuum as in the case of the scalar mode vacuum fluctuations. In this case, the amplitude is uniquely determined by the Hubble parameter $H$ during inflation, namely, it is characterized by the tensor power spectrum proportional to $H^2(t_k)$ where $t_k$ is the time at which the mode with comoving wavenumber $k$ left the Hubble horizon, and the bispectum is known to be small. 

However, there have been several proposals that may alter this prediction. One of such is the case of modified gravity, e.g., Horndeski theory, massive gravity, etc. 
Furthermore, it has been suggested that the quantum nature of the graviton state may be observationally detected if gravitons were not in the adiabatic 
vacuum~\cite{Kanno:2018cuk,Kanno:2019gqw}. This motivates us to study possible non-Gaussianities for non-adiabatic initial states.
In this paper, without specifying a particular model, we consider a vacuum state that deviates from the adiabatic one. More precisely, assuming that the inflationary stage can be approximated by a pure de Sitter space (or it can indeed be pure de Sitter in the case of false vacuum inflation~\cite{Copeland:1994vg} or the false vacuum stage prior to vacuum tunneling in open inflation~\cite{Sasaki:1993ha,Bucher:1994gb}), we consider the general $\alpha$-vacuum for the tensor perturbation, and compute the spectrum and bispectrum without assuming that the deviation from the BD vacuum is small. 

The $\alpha$-vacuum is known to be the general de Sitter invariant vacuum, specified by a real parameter $\alpha$ and a phase $\phi$~\cite{Allen:1985ux}. The case $\alpha=0$ is called the Bunch-Davies (BD) vacuum and it corresponds to the adiabatic vacuum in the general slow-roll inflation. A non-zero value of $\alpha$ means a deviation from the adiabatic vacuum, independent of the phase $\phi$.

The 3-point correlation function for primordial tensor fluctuations in the BD vacuum is first calculated in~\cite{Maldacena:2002vr}, and an extensive discussion on the possible forms of tensor non-Gaussianities due to the cubic Weyl terms are given in~\cite{Maldacena:2011nz, Soda:2011am, Shiraishi:2011st}.
The tensor bispectrum in the $\alpha$-vacuum are calculated for squeezed configurations  in~\cite{Kundu:2013gha}. In~\cite{Ashoorioon:2018sqb}, both squeezed and folded configurations of the bispectrum are considered with the assumption that there was a transition from the BD vacuum to an $\alpha$-vacuum at an early epoch. In the case of the scalar curvature perturbation, the 3-point function for the general $\alpha$-vacuum was derived in \cite{Xue:2008mk}.

In the case of the conventional slow-roll inflation, deviations from the adiabatic vacuum give rise to extra energy density carried by the field, and the condition that this extra energy density be smaller than the vacuum energy density responsible for inflation places a severe constraint on these deviations~\cite{Tanaka:2000jw}. Thus, in particular, in many of the previous studies on the tensor perturbation in the $\alpha$-vacuum, only small deviations from the BD vacuum were discussed~\cite{Akama:2020jko}. 
However, thus obtained constraints from the backreation arguments depend not only on the energy scale of inflation but also on the wavenumbers under consideration. Furthermore, in the case of false vacuum inflation when the spacetime is pure de Sitter, it seems there is no strong reason why the state cannot be in a non-trivial ($\alpha\neq0 $) $\alpha$-vacuum, as it respects the full de Sitter invariance. At the least, it is worth presenting the explicit form of the tensor bispectrum for the general $\alpha$-vacuum without any approximations.

The paper is organized as follows. In section 2, We briefly review the quantization of the tensor perturbation in de Sitter space, describe the $\alpha$-vacuum, and present the action up through cubic self-interactions. For the cubic self-interactions, we focus on the case of general relativity as its presence seems robust independent of models of interest.
In section 3, we compute the 2-point and 3-point functions in momentum space, which correspond to the spectrum and bispectrum, respectively, for the general $\alpha$-vacuum.
We find that the resulting bispectrum can become arbitrarily large, while the power spectrum is kept small, or even significantly suppressed for a particular choice of the phase $\phi$.
As particular cases of interest, we then take the squeezed and folded limits.
The result for the squeezed configuration is in agreement with \cite{Kundu:2013gha}.
Section 4 is devoted to conclusions and discussion. Some technical details are deferred to Appendix A.

\section{Graviton in de Sitter Space}
\label{section2}

The graviton in a spatially flat expanding background may be represented by the tensor mode perturbation in the three-dimensional metric,
\begin{eqnarray}
ds^{2}=a^{2}\left(\eta\right)\left[-d\eta^{2}+\left(\delta_{ij}+\gamma_{ij}\right)dx^{i}dx^{j}\right]\,,
\end{eqnarray}
where $\eta$ is the conformal time and the metric perturbation $\gamma_{ij}$ satisfies
the transvers traceless condition $\gamma_{ij}{}^{,j}=\gamma^{i}{}_{i}=0$. 
Here and in what follows the spatial indices $i,j,k,\cdots$ are raised and lowered by 
$\delta^{ij}$ and $\delta_{k\ell}$. 

We focus on general relativity. The Einstein-Hilbert action up through the third order in $h_{ij}\equiv \frac{M_{\rm pl}}{2}\gamma_{ij}$ is given by
\begin{eqnarray}
S=\frac{M_{\rm pl}^{2}}{2}\int d^{4}x\sqrt{-g}\,R
=S_{2}+S_{3}\,,
\end{eqnarray}
where $M_{\rm pl}^2=1/(8\pi G)$ and
\begin{eqnarray}
S_{2}&=&\frac{1}{2}\int d^{4}x\,a^{2}\left[
h^{ij\,\prime}\,h^{\prime}_{ij}-h^{ij,k}\,h_{ij,k}\right]\,,
\label{action2}\\
S_{3}&=&\frac{1}{M_{\rm pl}}
\int d^{4}x\,a^{2}\left[h^{ij}\,h_{k\ell,i}\,h^{k\ell}{}_{,j}-2h_{ij}\,h^{ik}{}_{,\ell}\,h^{j\ell}{}_{,k}
\right]\,.
\label{action3}
\end{eqnarray}
Here, a prime denotes the derivative with respect to the conformal time. 

\subsection{Free graviton in de Sitter space}
At quadratic order, we expand $h_{ij}(\eta,x^{i})$ in the Fourier modes,
\begin{eqnarray}
h_{ij}(\eta, x^i) = \int\frac{d^{3}k}{(2\pi)^{3/2}}\sum_{A}h^A_{\bm{k}}(\eta)\,e^{i {\bm k} \cdot {\bm x}} \ e_{ij}^A(\bm k)  \,,
\label{fourier}
\end{eqnarray}
where $e^A_{ij}({\bm k})$ is the polarization tensor for the $\bm k$ mode, normalized as $e^{ijA} e^B_{ij}{} =\delta^{AB}$ with $A,B=+,\times$. Then the quadratic action, (\ref{action2}), is rewritten as
\begin{eqnarray}
S_{2}=\frac{1}{2}\int d\eta\,d^{3}k\,\sum_{A}a^{2}\left[\,
|h^{A\,\prime}_{\bm k}(\eta)|^{2}-k^{2}\,|h_{\bm k}^{A}(\eta)|^{2}\,\right]\,,
\end{eqnarray}
where $k=|\bm k|$. The tensor mode $h^A_{\bm k}(\eta)$ satisfies
\begin{eqnarray}
h_{\bm k}^{A\,\prime\prime}+2\frac{a'}{a}h_{\bm k}^{A\,\prime} +k^2h^A_{\bm k}=0\,.
\label{eom}
\end{eqnarray}

Quantizing the above, the Fourier mode $h^A_{\bm k}(\eta)$ is promoted to an operator, and it may be expressed in terms of the creation and annihilation operators as
\begin{eqnarray}
h^A_{\bm k}(\eta)=b^A_{\bm k}\,u_k(\eta)+b_{-\bm k}^{A \dag}\,u_k^{*}(\eta)\,,
\label{operator1}
\end{eqnarray}
where
$
\left[b^A_{\bm k} , b_{\bm p}^{B\dag} \right]= \delta^{AB} \delta(\bm k-\bm p)
$, and the canonical commutation relation implies the Klein-Gordon normalization condition on 
the mode function $u_k(\eta)$ called the positive frequency function,
\begin{eqnarray}
u_k u_k^{*}{}'-u_k'u_k^*=\frac{i}{a^2}\,.
\end{eqnarray}

In the case of de Sitter space, the scale factor is given by $a(\eta)=-1/({H\eta})$ 
where $-\infty<\eta<0$. The de Sitter space is symmetric under SO(4,1) transformations. It is known that there are infinitely many vacua that respect the de Sitter symmetry, called the $\alpha$-vacua. Among them, a unique vacuum that corresponds to the adiabatic vacuum in an expanding universe is the BD vacuum. Assuming $b^A_{\bm k}$ is the operator that annihilates the BD vacuum,
$
b^A_{\bm k}\,|0_{\bm k} \rangle_{\rm BD} =0
$, the corresponding mode function $u_k$ is given by
\begin{eqnarray}
u_{k}(\eta)=\frac{H}{\sqrt{2k^{3}}}\left(1+ik\eta\right)e^{-ik\eta}
\,.
\label{positivefreq}
\end{eqnarray}

For an $\alpha$-vacuum, we consider another expansion of $h^A_{\bm k}(\eta)$,
\begin{eqnarray}
h^A_{\bm k}(\eta)=c^A_{\bm k}\,v_k(\eta)+c_{-\bm k}^{A \dag}\,v_k^{*}(\eta)\,,
\label{operator2}
\end{eqnarray}
where
$
\left[c^A_{\bm k} , c_{\bm p}^{B\dag} \right]= \delta^{AB} \delta_{\bm k,\bm p}
$, and $c^A_{\bm k}$ annihilates the $\alpha$-vacuum,
$
c^A_{\bm k}\,|0_{\bm k} \rangle_{\alpha} =0
$.
The mode functions and the creation and annihilation operators for the $\alpha$-vacuum are related to those for the BD vacuum by the Bogoliubov transformation,
\begin{eqnarray}
v_{k}(\eta)&=&\cosh\alpha\,u_{k}(\eta)+e^{{i\phi}}\sinh\alpha\,u^{*}_{k}(\eta)\,,
\label{bogoliubov1}\\
c^A_{\bm k}&=&\cosh\alpha\,b^A_{\bm k}-e^{{-i\phi}}\sinh\alpha\,b_{-\bm k}^{A \dag}\,,
\label{bogoliubov2}
\end{eqnarray}
where and below we assume $0\leq\alpha<\infty$ and $0\leq\phi<2\pi$ without loss of generality.

The $\alpha$-vacuum is a squeezed state from the point of view of the BD vacuum, and it may also be regarded as an entangled state. Applying Eq.~(\ref{bogoliubov2}) to $c^A_{\bm k}\,|0_{\bm k} \rangle_{\alpha} =0$ gives
\begin{eqnarray}
\left(\cosh\alpha\,b^A_{\bm k}-e^{{-i\phi}}\sinh\alpha\,b_{-\bm k}^{A \dag}\right)|0_{\bm k} \rangle_{\alpha}=0\,.
\end{eqnarray}
This leads to the expression of the $\alpha$-vacuum in terms of the BD vacuum as
\begin{eqnarray}
|0_{\bm k} \rangle_{\alpha}=N_k\exp\left[e^{i\phi}\tanh\alpha\, b^{A\dag}_{\bm k}\,b^{A\dag}_{-\bm k}\right]|0_{\bm k} \rangle_{\rm BD}|0_{-\bm k} \rangle_{\rm BD}\,,
\end{eqnarray}
where $N_k$ is a normalization constant. If we expand the exponential function in Taylor series, we find
\begin{eqnarray}
|0_{\bm k}\rangle_{\alpha}
&=&  N_k \Bigl(\, |0_{\bm k}\rangle_{\rm BD}\otimes |0_{-{\bm k}}\rangle_{\rm BD}
+e^{i\phi} \tanh \alpha\,|1_{\bm k}\rangle_{\rm BD}\otimes |1_{-{\bm k}}\rangle_{\rm BD} 
+\cdots 
\nonumber\\
&&\qquad
+e^{in\phi}\tanh^n \alpha\,|n_{\bf k}\rangle_{\rm BD}\otimes |n_{-{\bf k}}\rangle_{\rm BD}\Bigr) \,. 
\label{two-mode}
\end{eqnarray}
This is a two-mode squeezed state which consists of an infinite number of entangled particles. 
In particular, in the highly squeezing limit $\alpha\rightarrow\infty$, the $\alpha$-vacuum becomes the maximally entangled state from the point of view of the BD vacuum. 

Before closing this subsection, let us mention the issue of backreaction. 
It has been argued that any state that substantially deviates from the adiabatic vacuum would cause a backreaction problem because there would be a large number of excited particles with respect to the adiabatic vacuum (BD vacuum in the case of de Sitter space) as given in (\ref{two-mode}), whose energy density would dominate the universe and nullify inflation~\cite{Tanaka:2000jw}. If we apply this picture to our case, any non-trivial $\alpha$-vacuum would not be allowed as the parameter $\alpha$ is independent of the comoving momentum $k$. 

In this paper, as we mentioned briefly in the introduction, we take a more flexible point of view. In the case of conventional slow-roll inflation, the adiabatic vacuum is indeed the unique, physically most motivated vacuum. Hence we consider the $k$-independence as an approximation, and assume it applies only to momenta satisfying $k<k_{\rm max}$ for a certain maximum $k_{\rm max}$ as well as to the range of time $\eta_{i}<\eta$ for a certain initial time $\eta_i$, so that there would arise no backreaction problem. In addition, to simply the computations we assume $k|\eta_i|\gg1$ for the range of momenta of interest so that we may ignore the initial time dependence.
On the other hand, in the case of false vacuum inflation where the spacetime is exactly de Sitter, since all $\alpha$-vacua are de Sitter invariant,
we accept the possibility of an $\alpha$-vacuum as it is. In either case, we proceed to computing the leading order graviton non-Gaussianity in the $\alpha$-vaccum, 
hoping that it may shed some light on the initial state of the universe.

\subsection{Interaction picture in the in-in formalism}

In terms of the Fourier mode operators $h^A_{\bm k}(\eta)$, the cubic action Eq.~(\ref{action3}) 
is rewritten as
\begin{eqnarray}
&&S_{3}\equiv\int d\eta\, L_3\,;\nonumber\\
&&L_3=-\frac{a^2}{M_{\rm pl}(2\pi)^{3/2}}\sum_{A_{1}A_{2}A_{3}}
\int d^{3}p_{1}\int d^{3}p_{2}\int d^{3}p_{3}\,\delta\left({\bm p}_{1}+{\bm p}_{2}+{\bm p}_{3}
\right)h_{{\bm p}_{1}}^{A_{1}}(\eta)\,h_{{\bm p}_{2}}^{A_{2}}(\eta)\,
h_{{\bm p}_{3}}^{A_{3}}(\eta)
\nonumber\\
&&\qquad\times\left[
\,p_{2}^{i}\,p_{3}^j\,e_{ij}^{A_{1}}
\left({\bm p}_{1}\right)\,e^{A_{2}}_{k\ell}\left({\bm p}_{2}\right)e^{k\ell A_{3}}\left({\bm p}_{3}\right)
-2p_{2}^{\ell}\,p_{3}^{k}\,e^{ijA_{1}}
\left({\bm p}_{1}\right)\,e_{ik}^{A_{2}}\left({\bm p}_{2}\right)\,e_{j\ell}^{A_{3}}\left({\bm k}_{3}\right)\right]
\,.
\label{s3}
\end{eqnarray}
As usual, to take this self-interactions into account, we resort to the interaction picture, where the states and operators are expressed in terms of the free field operators with the interaction Hamiltonian $H_{I}=H-H_0$ being used to evolve the states and operators. Here $H$ and $H_0$ are the full and free Hamiltonians, respectively.
In the in-in formalism with which one computes the expectation value of an operator $Q$ at time $t$, we have
\begin{eqnarray}
\langle \psi_H|Q_H(t)|\psi_H\rangle
=\langle \psi|\bar T(e^{i\int_{-\infty}^t dt'\,H_{I}(t')})Q(t) T(e^{-i\int_{-\infty}^t dt'\,H_I(t')})|\psi\rangle\,,
\label{in-in}
\end{eqnarray}
where $|\psi_H\rangle$ ($Q_H$) is the state (operator) in the Heisenberg picture and $|\psi\rangle$ ($Q$) is the corresponding free state (operator) ,
and $T$ ($\bar T$) denotes the time-ordering (anti-time-ordering).

In our case, we set $H_0$ to be the one given by $L_2$ and 
$H_I=-L_3$, replace $t$ with the conformal time $\eta$, and consider the 2-point and 3-point functions,
\begin{eqnarray}
{}_{\alpha}\langle 0|\,\gamma_{{\bm k}_{1}}^{A_{1}}(\eta)\,\gamma_{{\bm k}_{2}}^{A_{2}}(\eta)\,|0\rangle_{\alpha}\,,
\quad
{}_\alpha\langle 0|\,
\gamma_{{\bm k}_{1}}^{A_{1}}(\eta)\,\gamma_{{\bm k}_{2}}^{A_{2}}(\eta)\,\gamma_{{\bm k}_{3}}^{A_{3}}(\eta)
\,|0\rangle_\alpha\,.
\label{23pointfcns}
\end{eqnarray}

\section{Two-point and Three-point Functions} 
\label{sec:2+3}

Following the in-in formalism presented in the previous section, we now compute the two-point and three-point functions at their leading orders.

\subsection{Two-point function}
The two-point function at leading order is simply given by its free field version. It is straightforward to obtain it by using Eqs.~(\ref{positivefreq}) and (\ref{bogoliubov1}).
Focusing on its behavior at late times ($\eta\rightarrow 0$), we obtain
\begin{eqnarray}
{}_{\alpha}\langle 0|\,\gamma_{{\bm k}_{1}}^{A_{1}}(\eta)\,\gamma_{{\bm k}_{2}}^{A_{2}}(\eta)\,|0\rangle_{\alpha}
=\delta\left({\bm k}_{1}+{\bm k}_{2}\right)\frac{\delta_{A_{1},A_{2}}}{2}P_T(k)\,,
\end{eqnarray}
where $k=|k_{1}|=|k_{2}|$, and $P_T(k)$ is the power spectrum given by
\begin{eqnarray}
P_T(k)=\frac{4}{k^3}\left(\frac{H}{M_{\rm pl}}\right)^2
\left(\cosh 2\alpha+\cos\phi\sinh 2\alpha\right)\,.
\label{spectrum}
\end{eqnarray}
The spectrum for the BD vacuum is realized when $\alpha=0$. Thus the $\alpha$-vacuum spectrum is generally enhanced by a factor $e^{2\alpha}$ for $\alpha\gg1$.

An intriguing fact is that, in the special case of $\phi=\pi$, the spectrum is exponentially suppressed like $e^{-2\alpha}$ for $\alpha\gg1$, instead of being enhanced.
This implies that the two-point function can be made arbitrarily small in principle. One might wonder how this could be consistently realized under the uncertainty principle, or under the Klein-Gordon normalization of the mode functions. The point is that although the $\alpha$-vacuum mode function $v_k$ is suppressed by $e^{-\alpha}$ at leading order in $|k\eta|$, its higher order terms are actually enhanced by $e^{\alpha}$. This leads to the enhancement by $e^\alpha$ of the component of $v_k'$ that contributes to the Klein-Gordon normalization. In other words, the canonical conjugate of $h_{ij}$, i.e., $a^2h_{ij}'$ is the one that may be regarded a frozen on superhorizon scales in the large $\alpha$ limit, $e^{2\alpha}|k\eta|\gg 1$. Possible effects of this on the arguments about classicalization and decoherence on superhorizon scales is an interesting issue, but it is out of the scope of the present paper.

\subsection{Three-point function}

The three-point function to its leading order is given by expanding the in-in formula to the first order in the interaction Hamiltonian,  
\begin{eqnarray}
&&{}_\alpha\langle 0_H|\,
\gamma_{{\bm k}_{1}}^{A_{1}}(\eta)\,\gamma_{{\bm k}_{2}}^{A_{2}}(\eta)\,\gamma_{{\bm k}_{3}}^{A_{3}}(\eta)
\,|0_H\rangle_\alpha\,
\nonumber\\
&&\qquad=-i\,{}_\alpha\langle 0|\left[
\gamma_{{\bm k}_{1}}^{A_{1}}(\eta)\,\gamma_{{\bm k}_{2}}^{A_{2}}(\eta)\,\gamma_{{\bm k}_{3}}^{A_{3}}(\eta),\int^\eta_{-\infty}d\eta_1\,H_{\rm int}(\eta_1)\right]
|0\rangle_\alpha\,.
\end{eqnarray}

Using Eqs.~(\ref{positivefreq}) and (\ref{bogoliubov1}) again, we obtain the three-point function in late times ($\eta\to0$) as
\begin{eqnarray}
{}_{\alpha}\langle 0|\,\gamma_{{\bm k}_{1}}^{A_{1}}(\eta)\,\gamma_{{\bm k}_{2}}^{A_{2}}(\eta)\,\gamma_{{\bm k}_{3}}^{A_{3}}(\eta)\,|0\rangle_{\alpha}
=\frac{1}{(2\pi)^{3/2}}\delta\left({\bm k}_{1}+{\bm k}_{2}+{\bm k}_{3}\right)
B({\bm k_1},{\bm k_2},{\bm k_3};A_1,A_2,A_3)\,,
\label{3pt1}
\end{eqnarray}
where $B$ is the bispectrum given by
\begin{eqnarray}
&&B({\bm k_1},{\bm k_2},{\bm k_3};A_1,A_2,A_3)=
-\frac{2}{(k_{1}k_{2}k_{3})^{3}}\left(\frac{H}{M_{\rm pl}}\right)^{4}
\Biggl[P({\bm k}_1,A_1\,\vert\,{\bm k}_2, {\bm k}_3; A_2, A_3)\,k_2k_3
\nonumber\\
&&\qquad\times\bigg\{\Bigl(\cosh^{6}\alpha-\sinh^{6}\alpha
+\frac{1}{4}\left(\cos\phi\sinh 4\alpha+\cos2\phi\sinh^{2}2\alpha\right)
\Bigr)
\nonumber\\
&&\qquad\quad\times
\left(k_{t}
-\frac{k_{1}k_{2}+k_{2}k_{3}+k_{3}k_{1}}{k_{t}}
-\frac{k_1k_2k_3}{k_{t}^2}
\right)
\nonumber\\
&&\left.~\qquad+\frac{1}{4}\Bigl(
\sinh^{2}2\alpha\cosh^{2}\alpha+\cos2\phi\sinh^{2}2\alpha\sinh^{2}\alpha
+\cos\phi\sinh 4\alpha\cosh^{2}\alpha\Bigr)\right.\nonumber\\
&&\left.\qquad\quad\times
\left(k_t'
-\frac{k_{1}k_{2}-k_{2}k_{3}-k_{3}k_{1}}{k_t'}
+\frac{k_1k_2k_3}{(k_t')^{2}}
\right)
\right.\nonumber\\
&&\left.~\qquad+\frac{1}{4}\Bigl(
\sinh^{2}2\alpha\sinh^{2}\alpha+\cos2\phi\sinh^{2}2\alpha\cosh^{2}\alpha
+\cos\phi\sinh 4\alpha\sinh^{2}\alpha\Bigr)\right.\nonumber\\
&&\left.\qquad\quad\times
\left(k_t''
-\frac{-k_{1}k_{2}+k_{2}k_{3}-k_{3}k_{1}}{k_t''}
-\frac{k_1k_2k_3}{(k_t'')^{2}}
\right)
\bigg\}\,+\,(k_{1},k_{2},k_{3}:{\rm cyclic})\right]\,,
\label{bispectrum}
\end{eqnarray}
where $P$ is the factor involving the polarization,
\begin{eqnarray}
&&P({\bm k}_1,A_1\,\vert\,{\bm k}_2, {\bm k}_3; A_2, A_3)
\nonumber\\
&&\qquad\equiv\frac{{k}_{2}^{i}\,{k}_{3}^{j}\,e_{ij}^{A_{1}}({\bm k}_{1})\,
e^{A_{2}}_{k\ell}({\bm k}_{2})e^{k\ell A_{3}}({\bm k}_{3})
-2\,{k}_{2}^{\ell}\,{k}_{3}^{k}\,e^{ijA_{1}}({\bm k}_{1})\,
e_{ik}^{A_{2}}({\bm k}_{2})\,e_{j\ell}^{A_{3}}({\bm k}_{3})}{k_2k_3}\,,
\label{Polarization}
\end{eqnarray}
and we have defined $k_{t}\equiv k_{1}+k_{2}+k_{3}$,
$k_{t}'\equiv k_1+k_2-k_3$, $k_{t}''\equiv k_1-k_2-k_3$, and it is understood that
$\bm{k}_1+\bm{k}_2+\bm{k}_3=0$ in the above.
The vertical bar separating $({\bm k}_1,A_1)$ and
$({\bm k}_2,{\bm k}_3;A_2,A_3)$ in the arguments of $P$ is inserted to remark the asymmetry between them.

As discussed in the previous subsection, the two-point function can be exponentially small for $\alpha\gg1$ if $\phi=\pi$. Therefore it is of interest to consider the three-point function in this case. Putting $\phi=\pi$ in the above, we find
\begin{eqnarray}
&&B({\bm k_1},{\bm k_2},{\bm k_3};A_1,A_2,A_3)|_{\phi=0}
=\frac{-2}{(k_{1}k_{2}k_{3})^{3}}
\left(\frac{H}{M_{\rm pl}}\right)^{4}
\Biggl[P({\bm k}_1,A_1\,\vert\,{\bm k}_2, {\bm k}_3; A_2, A_3)\,k_2k_3\nonumber\\
&&\times\bigg\{
\cosh2\alpha\Bigl(\cosh2\alpha-\cosh\alpha\sinh\alpha)\Bigr)
\Bigl(k_{t}
-\frac{k_{1}k_{2}+k_{2}k_{3}+k_{3}k_{1}}{k_{t}}
-\frac{k_1k_2k_3}{k_{t}^2}
\Bigr)
\nonumber\\
&&~\quad-\frac{e^{-\alpha}}{4}\sinh4\alpha\,\cosh\alpha
\left(k_t'
-\frac{k_{1}k_{2}-k_{2}k_{3}-k_{3}k_{1}}{k_t'}
+\frac{k_1k_2k_3}{(k_t')^{2}}
\right)
\nonumber\\
&&~\quad+\frac{e^{-\alpha}}{4}\sinh4\alpha\,\sinh\alpha
\left(k_t''
-\frac{-k_{1}k_{2}+k_{2}k_{3}-k_{3}k_{1}}{k_t''}
-\frac{k_1k_2k_3}{(k_t'')^{2}}
\right)
\bigg\}
\nonumber\\
&&\qquad\qquad+\,(k_{1},k_{2},k_{3}:{\rm cyclic})\Biggr]\,.
\label{phi=0}
\end{eqnarray}
The above implies that the bispectrum is enhanced by a factor $e^{4\alpha}$ for $\alpha\gg1$. Thus if the graviton was, or the graviton modes in a certain range of wavenumbers were in an $\alpha$-vacuum like state with $\alpha\gg1$ and $\phi=\pi$, the bispectrum may be exponentially enhanced relative to the amplitude of the two-point function. This suggests that the higher point functions are even more enhanced, and the perturbative expansion ceases to be valid in this particular case. Whether this is the case or not is an interesting mathematical question, and it would be intriguing if such a state is actually realized in some scenario of inflation. 

To detect the bispectrum in observation, it is customary to consider theoretical templates for the three limits in the momentum space configuration; the squeezed shape ($k_1\ll k_2\simeq k_3)$, the folded shape ($k_1=2k_2=2k_3)$, and the equilateral shape ($k_1=k_2=k_3$). They serve as theoretical templates.
In the current case, we immediately see that the bispectrum is enhanced for the squeezed and folded shapes. Therefore we focus on these two limits below.
Furthermore, it is customary to introduce the non-Gaussian parameter, say $f_{NL}$ for each of the three limits. However, as we are not sure if the same form of $f_{NL}$ used for the scalar type (i.e., curvature perturbation) non-Gaussianity could also be useful for the tensor type perturbation, because of the different $k$-dependence, we present it below just to get a sense of it.

Let us first consider the squeezed limit. Setting $k_1\ll k_2=k_3\equiv k$,
the leading order term is found as
\begin{eqnarray}
&&B({\bm k_1},{\bm k_2},{\bm k_3};A_1,A_2,A_3)|_{k_1\ll k_2=k_3=k}
=
\frac{-1}{k_{1}^4k^2}
\left(\frac{H}{M_{\rm pl}}\right)^{4}
P({\bm k}_1,A_1\,\vert\,{\bm k}_2, {\bm k}_3; A_2, A_3)
\nonumber\\
&&\qquad\qquad\times
\Bigl(
\sinh^{2}2\alpha\cosh^{2}\alpha+\cos2\phi\sinh^{2}2\alpha\sinh^{2}\alpha
+\cos\phi\sinh 4\alpha\cosh^{2}\alpha\Bigr)
\,.
\end{eqnarray}
Following the standard practice used for the curvature perturbation, we introduce the non-Gaussian parameter $f_{NL}^{\rm sq}$ for the squeezed limit defined by~\cite{Babich:2004gb}
\begin{eqnarray}
f_{NL}^{\rm sq}({\bm k_1},{\bm k_2},{\bm k_3};A_1,A_2,A_3)
\equiv
\frac{B({\bm k_1},{\bm k_2},{\bm k_3};A_1,A_2,A_3)}{P_T(k_1)P_T(k_2)+P_T(k_2)P_T(k_3)+P_T(k_3)P_T(k_1)}.
\label{fNLsqueezed}
\end{eqnarray}
Then we obtain
\begin{eqnarray}
&&f_{NL}^{\rm sq}({\bm k_1},{\bm k_2},{\bm k_3};A_1,A_2,A_3)
\simeq
\frac{-k}{32k_1}P({\bm k}_1,A_1\,\vert\,{\bm k}_2, {\bm k}_3; A_2, A_3)
\nonumber\\
&&\qquad\times
\frac{\sinh^{2}2\alpha\cosh^{2}\alpha
+\cos2\phi\sinh^{2}2\alpha\sinh^{2}\alpha
+\cos\phi\sinh 4\alpha\cosh^{2}\alpha}
{\left(\cosh 2\alpha+\cos\phi\sinh 2\alpha\right)^2}\,.
\end{eqnarray}
Thus $f_{NL}^{\rm sq}\propto k/k_1$. This implies $f_{NL}^{\rm sq}$ thus defined is strongly scale-dependent, unlike most cases of the scalar perturbation. In addition, $f_{NL}^{\rm sq}$ exponentially enhanced by a factor $\sim e^{2\alpha}$ if $\phi\neq\pi$, while by a factor $\sim e^{8\alpha}$ if $\phi=\pi$.

We note that if $f_{NL}^{\rm sq}$ is to be normalized with respect to the amplitude of the scalar curvature perturbation spectrum, it should be multiplied by $r^2$ where $r=P_T(k)/P_S(k)$ is the tensor-to-scalar ratio. As the difference in the $k$-dependence between $P_T(k)$ and $P_S(k)$ is small, the shape of $f_{NL}^{\rm sq}$ remains the same by the multiplication of $r^2$ at leading order. 

Next let us turn to the folded limit, $k_1=k_2+k_3$ with $k_2=k_3\equiv k$.
We immediately see that this limit is quadratic divergent because of the terms proportional to $(k_t')^{-2}$ and $(k_t'')^{-2}$ and their cyclic permutations. 
To clarify how this divergence would affect actual observables like the CMB anisotropy is beyond the scope of this paper. Nevertheless, since the tensor perturbation affects CMB only through its time derivative, we would expect that the observable effects are non-singular. In any case, let us evaluate the folded limit by introducing $\Delta k\equiv 2k-k_1$. At leading order, we obtain
\begin{eqnarray}
&&B({\bm k_1},{\bm k_2},{\bm k_3};A_1,A_2,A_3)|_{k_2=k_3=k, \Delta k=2k-k_1}
=\frac{1}{8k^4(\Delta k)^2}\left(\frac{H}{M_{\rm pl}}\right)^{4}
\\
&&\times\Biggl[P({\bm k}_1,A_1\,\vert\,{\bm k}_2, {\bm k}_3; A_2, A_3)
\Bigl(\sinh^{2}2\alpha(\sinh^{2}\alpha+\cos2\phi\cosh^{2}\alpha)
+\cos\phi\sinh 4\alpha\sinh^{2}\alpha\Bigr)
\nonumber\\
&&
\quad-2P({\bm k}_2,A_2\,\vert\,{\bm k}_3, {\bm k}_1; A_3, A_1)
\Bigl(\sinh^{2}2\alpha(\cosh^{2}\alpha+\cos2\phi\sinh^{2}\alpha)
+\cos\phi\sinh 4\alpha\cosh^{2}\alpha\Bigr)\,.
\Biggr]\nonumber
\label{folded}
\end{eqnarray}

Again, similar to the squeezed limit, by applying the standard convention of $f_{NL}^{\rm fo}$ used for the scalar type perturbation, defined by~\cite{Chen:2006nt}
\begin{eqnarray}
&&f_{NL}^{\rm fo}\equiv
B({\bm k_1},{\bm k_2},{\bm k_3};A_1,A_2,A_3)
\Bigl((P_T(k_1)P_T(k_2)+{\rm cyc.})
\nonumber\\
&&\qquad
+3(P_T(k_1)P_T(k_2)P_T(k_3))^{2/3}
-\left((P_T(k_1)P_T^2(k_2)P_T^3(k_3))^{1/3}+
{\rm perm.}\right)\Bigr)^{-1}\,,
\label{fNLfolded}
\end{eqnarray}
we find
\begin{eqnarray}
&&f_{NL}^{\rm fo}\simeq\frac{k^2}{32(\Delta k)^2}
\frac{1}{\left(\cosh 2\alpha+\cos\phi\sinh 2\alpha\right)^{2}}
\\
&&\times
\Biggl[P({\bm k}_1,A_1\,\vert\,{\bm k}_2, {\bm k}_3; A_2, A_3)
\Bigl(\sinh^{2}2\alpha(\sinh^{2}\alpha+\cos2\phi\cosh^{2}\alpha)
+\cos\phi\sinh 4\alpha\sinh^{2}\alpha\Bigr)
\nonumber\\
&&-2P({\bm k}_2,A_2\,\vert\,{\bm k}_3, {\bm k}_1; A_3, A_1)
\Bigl(\sinh^{2}2\alpha(\cosh^{2}\alpha+\cos2\phi\sinh^{2}\alpha)
+\cos\phi\sinh 4\alpha\cosh^{2}\alpha\Bigr)\,
\Biggr]\,.\nonumber
\end{eqnarray}
Thus, $f_{NL}^{\rm fo}$ is enhanced not only by a scale-dependent factor $k^2/(\Delta k)^2$ but also exponentially enhanced for $\alpha\gg1$; by a factor $\sim e^{2\alpha}$ if $\phi\neq\pi$ and  $\sim e^{8\alpha}$ if $\phi=\pi$. Also if we are to normalize $f_{NL}^{\rm fo}$ by the scalar curvature perturbation spectrum, it should be multiplied by
$r^2$.

In summary, both the squeezed and folded limits of non-Gaussianity are strongly scale-dependent, and are exponentially enhanced for $\alpha\gg1$. The enhancement factor is $\sim e^{2\alpha}$ if $\phi\neq\pi$, and $\sim e^{8\alpha}$ if $\phi=\pi$ for both limits. This result suggests that the parameter $\alpha$ may already be strongly constrained by the existing observational data such as the Planck CMB data~\cite{Planck:2019kim}.

\section{Conclusion} \label{section5}

In this paper, we computed the graviton bispectrum for the general $\alpha$-vacuum in de Sitter space, assuming the conventional cubic self-interactions in general relativity. 
The $\alpha$-vacuum is characterized by a real non-negative parameter $\alpha$ and a phase $\phi$. 

It was argued that the $\alpha$-vacuum is strongly constrained by the condition that its backreaction to the energy density of the universe would not jeopardize inflation~\cite{Tanaka:2000jw,Akama:2020jko}. This usually leads to the condition that the $\alpha$-vacuum parameter must be very small, $\alpha\ll1$, with its wavenumber range very narrow; $H<k/a\ll M_{\rm pl}$.
In this paper, we took a flexible point of view that the constraint could be considerably weakened if the energy scale of inflation was very low, or the universe was in a pure de Sitter phase where an $\alpha$-vacuum might be realized without backreaction. In the former case, recalling that the small backreaction requirement gives $e^{2\alpha}H^2/M_{\rm pl}^2\ll1$, $\alpha$ must satisfy the condition $e^{\alpha}\ll M_{\rm pl}/H$. 
If we consider inflation with the energy scale $\sim 10^3$GeV, for example, which is perfectly consistent with observational constraints as well as with the standard model of particle physics, we have $H^2/M_{\rm pl}^2\sim 10^{-64}$. This not only allows a large applicable range of wavenumbers but also implies a weak constraint, $e^\alpha\ll 10^{32}$ or $\alpha\ll 74$. The allowable range of $\alpha$ seems large enough to be interesting.

We found an intriguing result that the bispectrum may become very large in comparison with the spectrum for $\alpha\gg1$, and it becomes particularly enhanced if $\phi=\pi$. Namely, even if the tensor spectrum is too small to be detected by observation, there is a possibility that the bispectrum is large enough to be detected.

We have also computed the bispectrum in the squeezed limit ($k_1\ll k_2\simeq k_3$) and the folded limit ($k_1\gtrsim2k_2=2k_3$) for the general case of $\alpha$ and $\phi$. 
Introducing the non-Gaussian parameters $f_{NL}^{\rm sq}$ and $f_{NL}^{\rm fo}$ for these two limits, respectively, by copying their definitions in the case of the scalar bispectrum, we found that both non-Gaussian parameters are scale-dependent; $f_{NL}^{\rm sq}$ is proportional to $k/k_1$ where $k_1\ll k\equiv k_2\simeq k_3$, and $f_{NL}^{\rm fo}$ is proportional to $k^2/(\Delta k)^{2}$ where $\Delta k=k_1-k_2-k_3\equiv k_1-2k$. It is of interest to derive observational constraints on the $\alpha$-vacuum parameters from, say, the Planck CMB data~\cite{Planck:2019kim}. But it is out of the scope of the present paper.

Another, more fundamental issue is the validity of the perturbative expansion. In the case of $\alpha\gg1$, where the bispectrum is much larger than the spectrum, one may suspect that the higher order spectra could be even larger. This might be a sign of an unphysical property of the $\alpha$-vacuum. Nevertheless, at least from a phenomenological point of view, we believe that it is scientifically interesting enough to test the graviton $\alpha$-vacuum non-Gaussianities computed in this paper against observational data. We hope to come back to this in future work.

\section*{Acknowledgments}
We thank Takashi Hiramatsu, Jiro Soda and Takahiro Tanaka for useful discussion and comments.
This work was supported in part by the Japan Society for the Promotion of Science (JSPS) KAKENHI Grants No. JP22K03621 (SK) and Nos. 20H04727 and 20H05853 (MS). 

\appendix
\section{Three-point function at arbitrary time}
\label{appA}

In this appendix, we calculate the Fourier mode three-point function at arbitrary time. By expanding Eq.~(\ref{in-in}) to first order, we have
\begin{eqnarray}
{}_{\alpha}\langle 0|\,\gamma_{{\bm k}_{1}}^{A_{1}}(\eta)\,\gamma_{{\bm k}_{2}}^{A_{2}}(\eta)\,\gamma_{{\bm k}_{3}}^{A_{3}}(\eta)\,|0\rangle_{\alpha}
=-i\int^{\eta}_{-\infty}d\bar{\eta}\,
{}_{\alpha}\langle 0|
\left[\,\gamma_{{\bm k}_{1}}^{A_{1}}(\eta)\,\gamma_{{\bm k}_{2}}^{A_{2}}(\eta)\,\gamma_{{\bm k}_{3}}^{A_{3}}(\eta)\,,H_{\rm int}\left(\bar{\eta}\right)\right]
|0\rangle_{\alpha}\,,
\label{3pt1-app}
\end{eqnarray}
where $H_{\rm int}=-L_{3}$.
The right-hand side may be computed by noting that apart from the polarization dependence, it is proportional to $X(k_1,k_2,k_3)$ where
\begin{eqnarray}
&&X(k_1,k_2,k_3) 
\nonumber\\
&&~\equiv -i\Biggl\{
v_{k1}(\eta)v_{k2}(\eta)v_{k3}(\eta)
\left[\int_{-\infty}^{\eta}d\bar{\eta}\,a^{2}(\bar{\eta})\,
v^{*}_{k1}(\bar{\eta})v^{*}_{k2}(\bar{\eta})v^{*}_{k3}(\bar{\eta})
+(k_{1},k_{2},k_{3}:{\rm cyclic})\right]
\nonumber\\
&&~-v^{*}_{k1}(\eta)v^{*}_{k2}(\eta)v^{*}_{k3}(\eta)
\left[\int_{-\infty}^{\eta}d\bar{\eta}\,a^{2}(\bar{\eta})\,
v_{k1}(\bar{\eta})v_{k2}(\bar{\eta})v_{k3}(\bar{\eta})
+(k_{1},k_{2},k_{3}:{\rm cyclic})\right]\Biggr\}\,,
\label{3pt2}
\end{eqnarray}
where $v_k$ is the positive frequency mode function of the $\alpha$-vacuum.
The integrals may be expressed in terms of those involving the mode functions for the Bunch-Davies vacuum in Eq.~(\ref{positivefreq}). For example, we have
\begin{eqnarray}
&&\hspace{-1.0cm}\int_{-\infty}^{\eta}d\bar{\eta}\,a^{2}(\bar{\eta})
v_{p1}(\bar{\eta})v_{p2}(\bar{\eta})v_{p3}(\bar{\eta})\nonumber\\
&=&\cosh^{3}\alpha
\int_{-\infty}^{\eta}d\bar{\eta}\,a^{2}(\bar{\eta})\,u_{p1}(\bar{\eta})u_{p2}(\bar{\eta})u_{p3}(\bar{\eta})+e^{3i\phi}\sinh^{3}\alpha
\int_{-\infty}^{\eta}d\bar{\eta}\,a^{2}(\bar{\eta})\,u^{*}_{p1}(\bar{\eta})u^{*}_{p2}(\bar{\eta})u^{*}_{p3}(\bar{\eta})\nonumber\\
&&+e^{i\phi}\cosh^{2}\alpha\sinh\alpha
\int_{-\infty}^{\eta}d\bar{\eta}\,a^{2}(\bar{\eta})\,u_{p1}(\bar{\eta})u_{p2}(\bar{\eta})u^{*}_{p3}(\bar{\eta})\nonumber\\
&&+e^{2i\phi}\cosh\alpha\sinh^{2}\alpha
\int_{-\infty}^{\eta}d\bar{\eta}\,a^{2}(\bar{\eta})\,u_{p1}(\bar{\eta})u^{*}_{p2}(\bar{\eta})u^{*}_{p3}(\bar{\eta})\,,
\end{eqnarray}
where $u_k$ is the Bunch-Davies positive frequency mode function,
\begin{eqnarray}
u_k=\frac{H}{\sqrt{2k^3}}(1+ik\eta)e^{-ik\eta}\,.
\end{eqnarray}

To calculate the time integrals, we define
\begin{eqnarray}
I(p_1,p_2,p_3)\equiv
 \frac{\sqrt{8p_1p_2p_3}}{H}
\int_{-\infty}^{\eta}d\bar{\eta}\,a^{2}(\bar{\eta})\,u_{p1}(\bar{\eta})u_{p2}(\bar{\eta})u_{p3}(\bar{\eta})\,.
\end{eqnarray}
Then we have
\begin{eqnarray}
I(p_1,p_2,p_3)&=&\bigg\{\left(\frac{-1}{p_{1}p_{2}p_{3}\eta}+\frac{\eta}{p_1+p_2+p_3}\right)
\nonumber\\
&&
-i\left(\frac{1}{(p_1+p_2+p_3)^{2}}
+\frac{p_{1}p_{2}+p_{2}p_{3}+p_{3}p_{1}}{p_{1}p_{2}p_{3}(p_1+p_2+p_3)}\right)\bigg\}
e^{-i(p_1+p_2+p_3)\eta}
\,,
\end{eqnarray}
and we find the other integrals are given by
\begin{eqnarray}
&&\int_{-\infty}^{\eta}d\bar{\eta}\,a^{2}(\bar{\eta})\,u_{p1}(\bar{\eta})u_{p2}(\bar{\eta})u^{*}_{p3}(\bar{\eta})
= \frac{H}{\sqrt{8p_1p_2p_3}}I(p_1,p_2,-p_3)
\nonumber\\
&&\int_{-\infty}^{\eta}d\bar{\eta}\,a^{2}(\bar{\eta})\,u_{p1}(\bar{\eta})u^{*}_{p2}(\bar{\eta})u^{*}_{p3}(\bar{\eta})
=\frac{H}{\sqrt{8p_1p_2p_3}}I(p_1,-p_2,-p_3)
\nonumber\\
&&\int_{-\infty}^{\eta}d\bar{\eta}\,a^{2}(\bar{\eta})\,u^{*}_{p1}(\bar{\eta})u^{*}_{p2}(\bar{\eta})u^{*}_{p3}(\bar{\eta})
=\frac{H}{\sqrt{8p_1p_2p_3}}I(-p_1,-p_2,-p_3)\,.
\nonumber\\
\end{eqnarray}
Using the above results, we obtain 
\begin{eqnarray}
&& X({k}_1,{k}_2,{k}_3)=i\frac{H}{\sqrt{8k_1k_2k_3}}\Bigl\{
\cosh^3\alpha\,u_{k_1}^*u_{k_2}^*u_{k_3}^*
+e^{-3i\phi}\sinh^3\alpha\,u_{k_1}u_{k_2}u_{k_3}
\nonumber\\
&&\qquad +e^{-i\phi}\cosh^2\alpha\sinh\alpha\,u_{k_1}^*u_{k_2}^*u_{k_3}
+e^{-2i\phi}\cosh\alpha\sinh^2\alpha\,u_{k_1}^*u_{k_2}u_{k_3}
\Bigr\}\nonumber\\
&&\times\Bigl\{
\cosh^3\alpha\,I(k_1,k_2,k_3)+e^{3i\phi}\sinh^3\alpha\,I(-k_1,-k_2,-k_3)+e^{i\phi}\cosh^2\alpha\sinh\alpha\,I(k_1,k_2,-k_3)\nonumber\\
&&\qquad
+e^{2i\phi}\cosh\alpha\sinh^2\alpha\,I(k_1,-k_2,-k_3)
+(k_{1},k_{2},k_{3}:{\rm cyclic})
\Bigr\}\nonumber\\
&&-i\frac{H}{\sqrt{8k_1k_2k_3}}\Bigl\{
\cosh^3\alpha\,u_{k_1}u_{k_2}u_{k_3}
+e^{3i\phi}\sinh^3\alpha\,u_{k_1}^*u_{k_2}^*u_{k_3}^*
+e^{i\phi}\cosh^2\alpha\sinh\alpha\,u_{k_1}u_{k_2}u_{k_3}^*\nonumber\\
&&\qquad +e^{2i\phi}\cosh\alpha\sinh^2\alpha\,u_{k_1}u_{k_2}^*u_{k_3}^*+(k_{1},k_{2},k_{3}:{\rm cyclic})
\Bigr\}\nonumber\\
&&\times\Bigl\{
\cosh^3\alpha\,I^*(k_1,k_2,k_3)+e^{3i\phi}\sinh^3\alpha\,I^*(-k_1,-k_2,-k_3)+e^{i\phi}\cosh^2\alpha\sinh\alpha\,I^*(k_1,k_2,-k_3)\nonumber\\
&&\qquad
+e^{2i\phi}\cosh\alpha\sinh^2\alpha\,I^*(k_1,-k_2,-k_3)
+(k_{1},k_{2},k_{3}:{\rm cyclic})\Bigr\}\,.
\end{eqnarray}
If we take the $k\eta\rightarrow 0$ limit, we obtain Eq.~(\ref{bispectrum}).

\end{document}